\documentclass[11pt,twoside,onecolumn]{article}
\usepackage[]{latexsym}
\usepackage{epsfig}
\usepackage{amsmath,amssymb}
\RequirePackage[dvipsnames,usenames]{color}
\usepackage[colorlinks,hyperindex]{hyperref}

\definecolor{green1}{RGB}{0,128,0} 

\hypersetup{hidelinks,backref=true,pagebackref=true,hyperindex=true,colorlinks=true,breaklinks=true,urlcolor= blue}
\hypersetup{%
  colorlinks = true,
  linkcolor  = blue,
  citecolor = blue,
}
\usepackage{cite}
\newcommand{\be}{\begin{eqnarray}}
\newcommand{\ee}{\end{eqnarray}}

\newcommand{\rH}{r_{\rm H}}
\newcommand{\rh}{r_{\rm h}}

\title{\bf Black holes by gravitational decoupling}
\author{J.~Ovalle$^{ab}$\thanks{Corresponding author: jovalle@usb.ve}
$\,$, 
R.~Casadio$^{cd}$\thanks{casadio@bo.infn.it}
$\,$,
R.~da~Rocha$^{e}$\thanks{roldao.rocha@ufabc.edu.br} 
$\,$,
A.~Sotomayor$^{f}$\thanks{adrian.sotomayor@uantof.cl}
$\,$,
Z.~Stuchlik$^{a}$\thanks{zdenek.stuchlik@fpf.slu.cz}
\\
\null
\\
$^a${\em Institute of Physics and Research Centre of Theoretical Physics and Astrophysics,}
\\
{\em Faculty of Philosophy and Science, Silesian University in Opava}
\\
{\em CZ-746 01 Opava, Czech Republic}
\\
$^b${\em Departamento de F\'{\i}sica, Universidad Sim\'on Bol\'ivar,}
\\
{\em AP 89000, Caracas 1080A, Venezuela}
\\
$^c${\em Dipartimento di Fisica e Astronomia, Alma Mater Universit\`a di Bologna}
\\
{\em via Irnerio~46, 40126 Bologna, Italy}
\\
$^d${\em Istituto Nazionale di Fisica Nucleare, Sezione di Bologna, I.S.~FLAG}
\\
{\em  viale Berti~Pichat~6/2, 40127 Bologna, Italy}
\\
$^e${\em Centro de Matem\'atica, Computa\c c\~ao e Cogni\c c\~ao,}
\\
{\em Universidade Federal do ABC (UFABC)}
\\
{\em  09210-580, Santo Andr\'e, SP, Brazil.}
\\
$^f${\em Departamento de Matem\'aticas, Universidad de Antofagasta}
\\
{\em  Antofagasta, Chile}
}
\begin{document}
\maketitle
\begin{abstract}
We investigate how a spherically symmetric fluid modifies the Schwarzschild vacuum solution
when there is no exchange of energy-momentum between the fluid and the central source of the
Schwarzschild metric.
This system is described by means of the gravitational decoupling realised via the minimal geometric
deformation approach, which allows us to prove that the fluid must be anisotropic.
Several cases are then explicitly shown. 
\end{abstract}
%
%
%
%
%
%
%

\section{Introduction}
\setcounter{equation}{0}

The study of black holes represents one of the most active areas of gravitational physics,
from both a purely theoretical and the observational point of view.
The interest black holes generate is due not only to their exotic nature, but also because they
constitute ideal laboratories to study gravity in the strong field regime, and test general relativity
therein.
However, confronting theoretical predictions with observations is an arduous and complicated task.
A formidable step in this direction is the recent direct observation of black holes through the
detection of gravitational waves, which opens a new and promising era for gravitational
physics~\cite{ligo1,ligo2}.
\par
It is well known that general relativity predicts surprisingly simple solutions for black holes,
characterised at most by three fundamental parameters, namely the mass $M$,
angular momentum $J$ and charge $Q$~\cite{hawking}.
The original {\em no-hair\/} conjecture states that these solutions should not carry any other
charges~\cite{no-hair}.
Therefore, as the observations of systems containing black holes improve, the degree of
consistency of these observations with the predictions determined according to the
general relativistic solutions (with parameters $M$, $J$ and $Q$) will result
in a direct test of the validity of general relativity in the strong field regime.
There could in fact exist other charges associated with inner gauge symmetries
(and fields), and it is now known that black holes could have (soft) quantum
hair~\cite{Hawking:2016msc}.
The existence of new fundamental fields, which leave an imprint on the structure
of the black hole, thus leading to {\em hairy\/} black hole solutions, is precisely the scenario
under study in this paper.
\par
Possible conditions for circumventing the no-go theorem have been investigated for a long time
in different scenarios (see Refs.~\cite{thomas1,babi,thomas2,radu,marcelobh,thomas3,kanti1,adolfo1,adolfo2,adolfo3}
for some recent works and Refs.~\cite{galtsov1,kanti2,kanti3,galtsov2,mtz,konstantin} for earlier works).
In particular, a fundamental scalar field $\phi$ has been considered with great interest
(see Ref.~\cite{thomasBH} and references therein).
In this work, we will take a different and more general approach than most of the investigations
carried out so far and, instead of considering specific fundamental fields to generate hair in black hole
solutions, we shall just assume the presence of an additional completely generic source described by
a {\em conserved\/} energy-momentum tensor $\theta_{\mu\nu}$.
Of course, this $\theta_{\mu\nu}$ could account for one or more fundamental fields,
but the crucial property is that it gravitates but does not interact directly with the matter that
sources the (hairless) black hole solutions we start from.
This feature may seem fanciful, but can be fully justified, for instance, in the context of the
dark matter.
Achieving this level of generality in the classical scheme represented by general relativity
is a non-trivial task, and the gravitational decoupling by Minimal Geometric Deformation
(MGD-decoupling, henceforth) is precisely the method that was developed for this purpose
in Ref.~\cite{MGD-decoupling}.
\par
%
%
%
The MGD approach was originally proposed~\cite{jo1,jo2} in the context of the brane-world~\cite{lisa1,lisa2}
and extended to investigate new black hole solutions in Refs.~\cite{MGDextended1,MGDextended2}
(for some earlier works on the MGD, see for instance Refs.~\cite{jo6,jo8,jo9,jo10},
and Refs.~\cite{jo11,jo12,roldaoGL,rrplb,rr-glueball,rr-acustic,mgdanis,Luciano,camilo,ernesto,sharif1,sharif2,MGDBH,emgd,milko,tello,ernesto2}
for some recent applications).
The MGD-decoupling has a number of ingredients that make it particularly attractive in the search for new
spherically symmetric solutions of Einstein's field equations. The two main feature of this approach are the
following~\cite{MGD-decoupling}:
\begin{itemize}
\item 
{\it Extending simple solutions into more complex domains.} We can start from a simple
spherically symmetric gravitational source with energy-momentum tensor $\hat T_{\mu\nu}$
and add to it more and more complex gravitational sources, as long as the spherical symmetry is
preserved.
The starting source $\hat T_{\mu\nu}$ could be as simple as we wish, including the vacuum indeed,
to which we can add a first new source, say
\begin{equation}
\label{coupling}
\hat T_{\mu\nu}\mapsto \tilde T^{(1)}_{\mu\nu}=\hat T_{\mu\nu}+\alpha^{(1)}\,T^{(1)}_{\mu\nu}
\ ,
\end{equation}
where $\alpha^{(1)}$ is a constant that traces the effects of the new source $T^{(1)}_{\mu\nu}$.
We can then repeat the process with more sources, namely
\begin{equation}
\tilde T^{(1)}_{\mu\nu}\mapsto \tilde T^{(2)}_{\mu\nu}=\tilde T^{(1)}_{\mu\nu}+\alpha^{(2)}\,T^{(2)}_{\mu\nu}
\ ,
\end{equation}
and so on.
In this way, we can extend straightforward solutions of the Einstein equations associated
with the simplest gravitational source $\hat T_{\mu\nu}$ into the domain of more intricate
forms of gravitational sources $T_{\mu\nu}=\tilde T^{(n)}_{\mu\nu}$, step by step and systematically. 
We stress that this method works as long as the sources do not exchange energy-momentum among them,
namely
\begin{equation}
\nabla_{\mu}\hat T^{\mu\nu}
=
\nabla_{\mu} T^{(1)\mu\nu}
=
\ldots
=
\nabla_{\mu} T^{(n)\mu\nu}
=
0
\ ,
\label{nablas}
\end{equation}
which further clarifies that the constituents can only couple via gravity.
\par
\item {\it Deconstructing a complex gravitational source.} The converse of the above also works.
In order to find a solution to Einstein's equations with a complex spherically symmetric energy-momentum
tensor $T_{\mu\nu}$, we can split it into simpler components, say $\hat T_{\mu\nu}$ and $T^{(i)}_{\mu\nu}$,
provided they all satisfy Eq.~\eqref{nablas}, and solve Einstein's equations for each one of these parts.
Hence, we will have as many solutions as are the contributions $T^{(i)}_{\mu\nu}$ in the original
energy-momentum tensor.
Finally, by a straightforward combination of all these solutions, we will obtain the solution to the
Einstein equations associated with the original energy-momentum tensor $T_{\mu\nu}$.
\end{itemize}
Since Einstein's field equations are non-linear, the MGD-decoupling represents a breakthrough
in the search and analysis of solutions, especially when we deal with situations beyond trivial cases,
such as the interior of self-gravitating systems dominated by gravitational sources more
realistic than the ideal perfect fluid~\cite{lake2,visser2005}.
Of course, we remark that this decoupling occurs because of the spherical symmetry and
time-independence of the systems under investigation. 
%
%
%
\par
In analogy with the well-known electro-vacuum and scalar-vacuum, in this paper we will consider
a Schwarzschild black hole surrounded by a spherically symmetric ``tensor-vacuum'', represented
by the aforementioned $\theta_{\mu\nu}$.
Following the MGD-decoupling, we will separate the Einstein field equations in i) Einstein's equations
for the spherically symmetric vacuum and ii) a ``quasi-Einstein" system for the spherically symmetric
``tensor-vacuum''. 
The MGD procedure will then allow us to merge the Schwarzschild solution for i) with the solution
for the ``quasi-Einstein" system ii) into the solution for the complete system
``Schwarzschild + tensor-vacuum''.
Like the case of the electro-vacuum and (in some cases) scalar-vacuum, new black hole solutions
with additional parameters $q_i$ besides the mass $M$ can be obtained, each one associated with
a particular equation of state for the ``tensor-vacuum''.
Demanding the geometry is free of singularities and other pathologies, implies regularity conditions
which show that not all of these parameters $q_i$ can be independent.  
\par
The paper is organised as follows:
in Section~\ref{s2}, we first review the fundamentals of the MGD-decoupling applied
to a spherically symmetric system containing a perfect fluid and an additional source $\theta_{\mu\nu}$;
in Section~\ref{s4}, new hairy black holes solutions are found by assuming the perfect fluid
has sufficiently small support so that only $\theta_{\mu\nu}$ exists outside the horizon;
finally, we summarise our conclusions in Section~\ref{s6}.
\section{MGD decoupling for a perfect fluid}
\label{s2}
\setcounter{equation}{0}
\par
Let us start from the standard Einstein field equations
\begin{equation}
\label{corr2}
R_{\mu\nu}-\frac{1}{2}\,R\, g_{\mu\nu}
=
-k^2\,T^{\rm (tot)}_{\mu\nu}
\ ,
\end{equation}
and assume the total energy-momentum tensor contains two contributions, namely
\begin{equation}
\label{emt}
T^{\rm (tot)}_{\mu\nu}
=
T^{\rm (m)}_{\mu\nu}+\alpha\,\theta_{\mu\nu}
\ ,
\end{equation}
where
\begin{equation}
\label{perfect}
T^{\rm (m)}_{\mu \nu }=(\rho +p)\,u_{\mu }\,u_{\nu }-p\,g_{\mu \nu }
\end{equation}
is the 4-dimensional energy-momentum tensor of a perfect fluid with 4-velocity field $u^\mu$,
density $\rho$ and isotropic pressure $p$.
The term $\theta_{\mu\nu}$ in Eq.~(\ref{emt}) describes an additional source whose coupling
to gravity is proportional to the constant $\alpha$~\cite{Matt}.
This source may contain new fields, like scalar, vector and tensor fields,
and will in general produce anisotropies in self-gravitating systems.
In any case, since the Einstein tensor satisfies the Bianchi identity, the total source in Eq.~(\ref{emt})
must satisfy the conservation equation
\begin{equation}
\nabla_\mu\,T^{{\rm (tot)}{\mu\nu}}=0
\ .
\label{dT0}
\end{equation}
\par 
We next specialise to spherical symmetry and no time-dependence.
In Schwarzschild-like coordinates, a static spherically symmetric metric $g_{\mu\nu}$ reads 
\begin{equation}
ds^{2}
=
e^{\nu (r)}\,dt^{2}-e^{\lambda (r)}\,dr^{2}
-r^{2}\left( d\theta^{2}+\sin ^{2}\theta \,d\phi ^{2}\right)
\ ,
\label{metric}
\end{equation}
where $\nu =\nu (r)$ and $\lambda =\lambda (r)$ are functions of the areal
radius $r$ only, ranging from $r=0$ (the star center) to some $r=R$ (the
star surface), and the fluid 4-velocity is given by $u^{\mu }=e^{-\nu /2}\,\delta _{0}^{\mu }$
for $0\le r\le R$.
The metric~(\ref{metric}) must satisfy the Einstein equations~(\ref{corr2}),
which explicitly read
\begin{eqnarray}
\label{ec1}
k^2
\left(
\rho+\alpha\,\theta_0^{\ 0}
\right)
&\!\!=\!\!&
\frac 1{r^2}
-
e^{-\lambda }\left( \frac1{r^2}-\frac{\lambda'}r\right)\ ,
\\
\label{ec2}
k^2
\left(-p+\alpha\,\theta_1^{\ 1}\right)
&\!\!=\!\!&
\frac 1{r^2}
-
e^{-\lambda }\left( \frac 1{r^2}+\frac{\nu'}r\right)\ ,
\\
\label{ec3}
k^2
\left(-p+\alpha\,\theta_2^{\ 2}\right)
&\!\!=\!\!&
\frac {e^{-\lambda }}{4}
\left( -2\,\nu''-\nu'^2+\lambda'\,\nu'
-2\,\frac{\nu'-\lambda'}r\right)
\ ,
\end{eqnarray}
where $f'\equiv \partial_r f$ and spherical symmetry implies that $\theta_3^{\ 3}=\theta_2^{\ 2}$.
The conservation equation~(\ref{dT0}) is a linear combination of Eqs.~(\ref{ec1})-(\ref{ec3}), and yields
\begin{equation}
\label{con1}
p'
+
\frac{\nu'}{2}\left(\rho+p\right)
-
\alpha\left(\theta_1^{\ 1}\right)'
+
\frac{\nu'}{2}\alpha\left(\theta_0^{\ 0}-\theta_1^{\ 1}\right)
+
\frac{2\,\alpha}{r}\left(\theta_2^{\ 2}-\theta_1^{\ 1}\right)
=
0
\ ,
\end{equation}
We then note the perfect fluid case is formally recovered for $\alpha\to 0$.
\par
The Eqs.~(\ref{ec1})-(\ref{ec3}) contain seven unknown functions, namely:
two physical variables, the density $\rho(r)$ and pressure $p(r)$;
two geometric functions, the temporal metric function $\nu(r)$
and the radial metric function $\lambda(r)$;
and three independent components of $\theta_{\mu\nu}$.
This system of equations is therefore indeterminate and we should emphasise that
the space-time geometry does not allow one to resolve for the gravitational source
$\{\rho, p, \theta_{\mu\nu}\}$ uniquely.
\par
In order to simplify the analysis, and by simple inspection, we can identify an effective density  
\begin{equation}
\tilde{\rho}
=
\rho
+\alpha\,\theta_0^{\ 0}
\ ,
\label{efecden}
\end{equation}
an effective radial pressure
\begin{equation}
\tilde{p}_{r}
=
p-\alpha\,\theta_1^{\ 1}
\ ,
\label{efecprera}
\end{equation}
and an effective tangential pressure
\begin{equation}
\tilde{p}_{t}
=
p-\alpha\,\theta_2^{\ 2}
\ .
\label{efecpretan}
\end{equation}
These definitions clearly illustrate that $\theta_{\mu\nu}$ could in general induce an
anisotropy,
\begin{equation}
\label{anisotropy}
\Pi
\equiv
\tilde{p}_{t}-\tilde{p}_{r}
=
\alpha\left(\theta_1^{\ 1}-\theta_2^{\ 2}\right)
\ ,
\end{equation}
inside a stellar distribution sourced by $T^{\rm (m)}_{\mu\nu}$ alone.
The system of Eqs.~(\ref{ec1})-(\ref{ec3}) may therefore be formally
treated as an anisotropic fluid~\cite{Luis,tiberiu}.
\par
The MGD-decoupling can now be applied to the case at hand by simply noting that the
energy-momentum tensor~\eqref{emt} is precisely of the form~\eqref{coupling}, with
$\hat T_{\mu\nu}=T^{\rm (m)}_{\mu\nu}$, $\alpha^{(1)}=\alpha$ and $T_{\mu\nu}^{(1)}=\theta_{\mu\nu}$.
The components of the diagonal metric $g_{\mu\nu}$ that solve the complete Einstein
equations~\eqref{corr2} and satisfy the MGD read~\cite{MGD-decoupling}
\begin{equation}
g_{\mu\nu}
=
\hat{g}_{\mu\nu}={g}_{\mu\nu}^{(1)}
\label{mgdi}
\end{equation}
for $\mu=\nu\neq 1$, and
\begin{equation}
g^{11}
=
\hat{g}^{11}+\alpha\,g^{(1)11}
\ ,
\label{mgd1}
\end{equation}
so that only the radial component is affected by the additional source $\theta_{\mu\nu}$.
This metric $g_{\mu\nu}$ is found by first solving the Einstein equations for the perfect fluid
source $T^{\rm (m)}_{\mu\nu}$,
\begin{equation}
\label{f1}
\hat{G}_{\mu\nu}
=
-k^2\,T^{\rm (m)}_{\mu\nu}
\ ,
\qquad
\qquad
\nabla_\mu T^{\rm (m)\mu\nu}=0
\ ,
\end{equation}
and then the remaining quasi-Einstein equations for the source $\theta_{\mu\nu}$,
namely
\begin{equation}
\tilde{G}_{\mu\nu}
=
-k^2\,{\theta}_{\mu\nu}
\ ,
\qquad\qquad
\nabla_\mu\theta^{\mu\nu}=0
\ ,
\label{MGD2}
\end{equation}
where the divergence-free quasi-Einstein tensor  
\begin{equation}
\tilde{G}_{\mu}^{\,\,\,\,\nu}
=
{G}_{\mu}^{\ \nu}+\Gamma_{\mu}^{\ \nu}
\ ,
\end{equation}
with $\Gamma_{\mu}^{\ \nu}$ a tensor that depends exclusively on ${g}_{\mu\nu}$ to ensure
the divergence-free condition.
For the spherically symmetric metric~\eqref{metric}, it reads
\begin{equation}
\label{Gamma}
\Gamma_{\mu}^{\ \nu}
=
\frac{1}{r^2}\left(\delta_\mu^{\ 0}\,\delta_0^{\ \nu}+\delta_\mu^{\ 1}\,\delta_1^{\ \nu}\right)
\ .
\end{equation}
\par
We can then proceed by considering a solution to the Eqs.~\eqref{f1} for a perfect fluid
[that is Eqs.~(\ref{ec1})-(\ref{con1}) with $\alpha=0$], which we can write as
\begin{equation}
ds^{2}
=
e^{\xi (r)}\,dt^{2}
-
\frac{dr^{2}}{\mu(r)}
-
r^{2}\left( d\theta^{2}+\sin ^{2}\theta \,d\phi ^{2}\right)
\ ,
\label{pfmetric}
\end{equation}
where 
\begin{equation}
\label{standardGR}
\mu(r)
\equiv
1-\frac{k^2}{r}\int_0^r x^2\,\rho(x)\, dx
=
1-\frac{2\,m(r)}{r}
\end{equation}
is the standard General Relativity expression containing the Misner-Sharp mass function $m(r)$.
The effects of the source $\theta_{\mu\nu}$ on the perfect fluid solution $\{\xi,\mu\,\rho,p\}$ can then
be encoded in the MGD undergone solely by the radial component of the perfect fluid geometry in Eq.~(\ref{pfmetric}).
Namely, the general solution is given by Eq.~\eqref{metric} with $\nu(r)=\xi(r)$ and
\begin{eqnarray}
\label{expectg}
e^{-\lambda(r)}
=
\mu(r)+\alpha\,f^{*}(r)
\ ,
\end{eqnarray}
where $f^*=f^*(r)$ is the MGD function to be determined from the quasi-Einstein Eqs.~\eqref{MGD2}, which
explicitly read
\begin{eqnarray}
\label{ec1d}
k^2\,\theta_0^{\ 0}
&\!\!=\!\!&
-\frac{f^{*}}{r^2}
-\frac{f^{*'}}{r}
\ ,
\\
\label{ec2d}
k^2\,\theta_1^{\ 1}
&\!\!=\!\!&
-f^{*}\left(\frac{1}{r^2}+\frac{\xi'}{r}\right)
\ ,
\\
\label{ec3d}
k^2\,\theta_2^{\ 2}
=
k^2\,\theta_3^{\ 3}
&\!\!=\!\!&
-\frac{f^{*}}{4}\left(2\,\xi''+\xi'^2+2\,\frac{\xi'}{r}\right)
-\frac{f^{*'}}{4}\left(\xi'+\frac{2}{r}\right)
\ .
\end{eqnarray}
We also notice that the conservation equations for the additional energy-momentum tensor,
$\nabla_\mu\theta^{\mu\nu}=0$, yield
\begin{equation}
\label{con1d}
\left(\theta_1^{\ 1}\right)'
-\frac{\xi'}{2}\left(\theta_0^{\ 0}-\theta_1^{\ 1}\right)
-\frac{2}{r}\left(\theta_2^{\ 2}-\theta_1^{\ 1}\right)
=
0
\ ,
\end{equation}
which does not depend on the MGD function $f^*$.
\par
In the next section, we shall solve the above equations starting from the simplest vacuum
solution given by the outer Schwarzschild metric, 
\begin{equation}
\label{Schw0}
ds^2
=
\left(1-\frac{2\,M}{r}\right)dt^2
-\left(1-\frac{2\,M}{r}\right)^{-1}dr^2
-d\Omega^2
\ ,
\end{equation}
therefore in a region of space where the perfect fluid $\rho$ and $p$ vanish.
\section{Black holes}
\label{s4}
\begin{figure}[t]
\center
\includegraphics[scale=0.30]{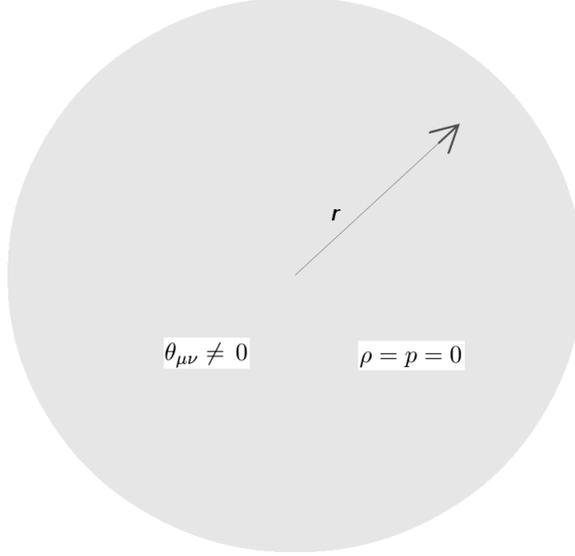}
\\
\caption{Spherically symmetric source $\theta_{\mu\nu}$ in the vacuum $\rho=p=0$.}
\label{ext}      
\end{figure}

When new paradigms beyond Einstein gravity are studied, the important question arises 
whether or not new black hole solutions exist. 
In order to address this point in general, we start from the results of the previous section and
determine the MGD function $f^{*}$ for the vacuum Schwarzschild solution~\eqref{Schw0}.
Figure~\ref{ext} schematically shows the kind of system we deal with.
The MGD metric will therefore read
\begin{equation}
\label{Schw}
ds^2
=
\left(1-\frac{2\,M}{r}\right)dt^2
-\frac{dr^2}{\strut\displaystyle{1-\frac{2\,M}{r}+\alpha\,f^{*}(r)}}
-r^2\,d\Omega^2
\ ,
\end{equation}
where the MGD function $f^{*}$ can be determined by imposing restrictions on the 
energy-momentum $\theta_{\mu\nu}$ to close the system of Eqs.~(\ref{ec1d})-(\ref{ec3d}).
\par
In the following we shall explore specific equations of state for $\theta_{\mu\nu}$ and
impose basic constraints on the causal structure of the resulting space-time in order to
have a well-defined horizon structure.
In particular, we recall that for the Schwarzschild metric~\eqref{Schw0}, the surface $\rH=2\,M$
is both a Killing horizon (determined by the condition $e^\nu=0$) and an outer marginally trapped
surface (the causal horizon, in brief, determined by the condition $e^{-\lambda}=0$).
For the MGD Schwarzschild metric~\eqref{Schw}, the component $g_{tt}=e^\nu$ is always
equal to the Schwarzschild form in Eq.~\eqref{Schw0} and can only vanish at $r=\rH$.
This means that $\rH=2\,M$ is still a Killing horizon (which can also become a real singularity).
However, the causal horizon is found at $r=\rh$ such that $g^{rr}(\rh)=e^{-\lambda}=0$, or
\begin{equation}
\label{newBH}
\rh
\left[1+\alpha\,f^{*}(\rh)\right]
=
2\,M
\ .
\end{equation}
We should therefore require that $\rh\ge 2\,M$, so that the surface $r=\rH$ is hidden behind
(or coincides with) the causal horizon.
Moreover, if $\rh>\rH$, the signature of the metric becomes $(++--)$ for $\rH<r<\rh$,
which one might want to discard as well, since not only the expansion of outgoing geodesics
vanishes for $r\to \rh^+$, but also ingoing geodesics never cross $r=\rh$: 
in this case the surface $r=\rh$ would act as a border of the outer space-time manifold.
To summarise, the MGD metric~\eqref{Schw} represents a proper black hole
only if the causal horizon coincides with the Killing horizon, that is when $\rh=\rH=2\,M$,
and this is therefore the condition we shall require in the following.
%
%
%
\subsection{Isotropic sector}
\par
Let us start by considering the case of isotropic pressure, so that 
\begin{equation}
\label{isotro}
\theta_1^{\ 1}
=\theta_2^{\ 2}
=\theta_3^{\ 3}
\ .
\end{equation}
Eqs.~(\ref{ec2d}) and (\ref{ec3d}) then yield a differential equation for the MGD function, namely
\begin{equation}
\label{giso}
f^{*'}\left(\xi'+\frac{2}{r}\right)
+f^{*}\left(2\,\xi''+\xi'^2-2\,\frac{\xi'}{r}-\frac{4}{r^2}\right)
=
0
\ ,
\end{equation}
whose general solution is given by
\begin{equation}
f^*(r)
=
\left(1-\frac{2\,M}{r}\right)
\left(\frac{r-M}{\ell_{\rm iso}}\right)^2
\ ,
\label{giso2}
\end{equation}
where $\ell_{\rm iso}$ is a constant with dimensions of a length. 
Hence, the MGD radial component for an isotropic deformation of the Schwarzschild exterior becomes
\begin{equation}
e^{-\lambda}
=
e^\xi
+
\alpha\,f^*
=
\left(1-\frac{2\,M}{r}\right)
\left[1
+
{\alpha}
\left(\frac{r-M}{\ell_{\rm iso}}\right)^2
\right]
\ ,
\end{equation}
which is clearly not asymptotically flat for $r\gg M$~\footnote{In fact, it approaches the radial component of the
de~Sitter metric for $r\sim\ell_{\rm iso}\gg M$.}.
We therefore conclude that the additional source $\theta_{\mu\nu}$ cannot contain an isotropic pressure
if we wish to preserve asymptotic flatness.
\subsection{Conformal sector}
\label{cs}
The energy-momentum tensor for a conformally symmetric source must be traceless. 
Since $\theta_2^{\ 2}=\theta_3^{\ 3}$, we therefore assume 
\begin{equation}
\label{ec3dc}
2\,\theta_2^{\ 2}
=
-\theta_0^{\ 0}-\theta_1^{\ 1}
\ ,
\end{equation}
so that the system~(\ref{ec1d})-(\ref{ec3d}) becomes
\begin{eqnarray}
\label{ec1dc}
-k^2\,\theta_0^{\ 0}
&\!\!=\!\!&
\frac{f^{*}}{r^2}
+\frac{f^{*'}}{r}
\\
\label{ec2dc}
-k^2\,\theta_1^{\ 1}
&\!\!=\!\!&
f^{*}\left(\frac{1}{r^2}+\frac{\xi'}{r}\right)
\ ,
\end{eqnarray}
where $f^{*}$ is again MGD function and $\xi$ the unperturbed Schwarzschild function.
From Eq.~\eqref{ec3dc}, we find the radial deformation must satisfy the differential equation
\begin{equation}
\label{fconf}
f^{*'}\left(\frac{\xi'}{2}+\frac{2}{r}\right)
+
f^{*}\left(\xi''+\frac{{\xi'}^2}{2}+2\,\frac{\xi'}{r}+\frac{2}{r^2}\right)
=
0
\ ,
\end{equation}
and it is important to highlight that the conservation equation~(\ref{con1d}) remains a linear
combination of the system~(\ref{ec1dc})-(\ref{ec3dc}).
The general solution for Eq.~(\ref{fconf}) is given by
\begin{equation}
\label{gconf}
f^{*}(r)
=
\frac{1-2\,M/r}{2\,r-{3\,M}}
\,\ell_{\rm c}
\ , 
\end{equation}
with $\ell_{\rm c}$ a constant with units of a length.
Thus the conformally deformed Schwarzschild exterior becomes
\begin{equation}
\label{confsol}
e^{-\lambda}
=
\left(1-\frac{2\,M}{r}\right)
\left(1+\frac{\ell}{2\,r-{3\,M}}\right)
\ ,
\end{equation}
where $\ell=\alpha\,\ell_{\rm c}$, and its behaviour for $r\gg M$ is given by
\begin{equation}
e^{-\lambda}
\simeq
1
-
\frac{4\,M-\ell}{2\,r}
\ .
\end{equation}
\par
The causal structure for this geometry is now more involved.
We still have the Killing horizon of the Schwarzschild metric at $\rH=2\,M$,
but $e^{-\lambda}$ diverges for
\begin{equation}
r_{\rm c}
=
\frac{3\,M}{2}
\ ,
\end{equation}
and there is a second zero of $e^{-\lambda}$ at
\begin{equation}
r_0
=
\frac{3\,M-\ell}{2}
=
r_{\rm c}
-\frac{\ell}{2}
\ .
\end{equation}
We can thus rewrite the radial metric component as 
\begin{equation}
\label{confsol2}
e^{-\lambda}
=
\left(1-\frac{\rH}{r}\right)
\left(\frac{r-r_0}{r-r_{\rm c}}\right)
\ .
\end{equation}
Note that $r_{\rm c}<\rH$ but, depending on the sign ands size of $\ell$, the second zero could occur
inside or outside the critical radius $r_{\rm c}$ and the Killing horizon $\rH$.
\par
In order to clarify the nature of the above solution, we compute explicitly the effective density 
\begin{equation}
\tilde{\rho}
=
\alpha\,\theta_0^{\ 0}
=
-\frac{\ell\,M}{4\,k^2\,(r-r_{\rm c})^2\,r^2}
\ ,
\label{efecdenC}
\end{equation}
the effective radial pressure
\begin{equation}
\tilde{p}_{r}
=
-\alpha\,\theta_1^{\ 1}
=
\frac{\ell}{2\,k^2\,(r-r_{\rm c})\,r^2}
\ ,
\label{efecpreraC}
\end{equation}
and the effective tangential pressure
\begin{equation}
\tilde{p}_{t}
=
-\alpha\,\theta_2^{\ 2}
=
\frac{\ell\,(r-M)}{4\,k^2\,(r-r_{\rm c})^2\,r^2}
\ .
\label{efecptanC}
\end{equation}
The anisotropy is thus given by
\begin{equation}
\Pi
=
\frac{\ell\,(3\,r-4\,M)}{k^2\,(2\,r-3\,M)^2\,r^2}
\ .
\end{equation}
The first thing we notice is that the density and pressures are regular on both $\rH$ and $r_0$,
but diverge at $r=r_{\rm c}<\rH$, which is therefore a real singularity, albeit hidden inside
the Killing horizon.
\par
We can then assume the black hole space-time is represented by the range $r>r_{\rm c}$, 
for which we must require that the region $r_{\rm c}<r<\rH$ has the proper signature,
as discussed previously.
This means that we must have $r_0\le r_{\rm c}$, or 
\begin{equation}
\label{cscondition}
\ell
>
0
\ ,
\end{equation}
with $\ell=0$ of course representing the pure Schwarzschild geometry.
We conclude that the conformal geometry in~\eqref{confsol} represents a black hole solution
with outer horizon $r_{\rm H}=2\,M$, and primary hairs represented by the parameter $\ell$,
which is constrained by the regularity condition~\eqref{cscondition}.
A solution similar to that in~(\ref{confsol}) was found in the context of the extra-dimensional
brane-world~\cite{cristiano}.
\subsection{Barotropic equation of state}
\label{pc}
If the additional source is a polytropic fluid, it should satisfy the equation of state
\begin{equation}
\label{polyt0}
\tilde p_r
=
K\,\tilde\rho^\Gamma
\ ,
\end{equation}
with $\Gamma=1+1/n$, where $n$ is the polytropic index and $K>0$ denotes a parameter which
contains the temperature implicitly and is governed by the thermal characteristics of a given polytrope.
For instance, the ultrarelativistic degenerate Fermi gas has polytropic index $n = 3$, while the non-relativistic
degenerate Fermi gas is found for $n = 3/2$.
(for more details, see for instance Refs.~\cite{willians,zdenek1,zdenek2,zdenek3,tooper}).
However, due to the unknown nature of the source $\theta_{\mu\nu}$, we will include the possibility
that $K<0$.
From Eqs.~\eqref{efecden} and \eqref{efecprera} with $\rho=p=0$, we then obtain
\begin{equation}
\label{polyt}
-\alpha\,\theta_1^{\ 1}
=
K
\left(\alpha\,\theta_0^{\ 0}\right)^\Gamma
\ .
\end{equation}
By using Eqs.~(\ref{ec1d}) and (\ref{ec2d}) in the expression~(\ref{polyt}) we obtain a first order
non-linear differential equation for the MGD function,  
\begin{equation}
\label{poly2}
{f^*}'
+
\frac{f^*}{r}
=
-\frac{1}{K^{1/\Gamma}}
\left(\frac{k^2\,r}{\alpha}\right)^{1-1/\Gamma}
\left(
\frac{f^*}{r-2\,M}
\right)^{1/\Gamma}
\ .
\end{equation}
We immediately notice that the right hand side is well-defined for a generic $\Gamma$ only provided $f^*/(r-2\,M)>0$.
\par
In order to proceed, we thus consider the simplest case $\Gamma=1$, so that Eq.~(\ref{polyt}) becomes
a barotropic equation of state.
This corresponds to an isothermal self-gravitating sphere of gas and is thus more appropriate
for our purpose.
It is worth mentioning that this self-gravitating sphere can also describe the collisionless
system of stars in a globular cluster.
The geometric deformation for $\Gamma=1$ and $r>2\,M$ simplifies to
\begin{equation}
f^{*}(r)
=
\left(1-\frac{2\,M}{r}\right)^{-1/K}
\left(\frac{\ell_p}{r}\right)^{1+1/K}
\ ,
\end{equation}
where $\ell_p>0$ is a length, and the MGD radial component of the metric reads
\begin{equation}
\label{poly}
e^{-\lambda}
=
\left(1-\frac{2\,M}{r}\right)
\left[1+\alpha\,\left(\frac{\ell_p}{r-2\,M}\right)^{1+1/K}\right]
\ ,
\end{equation}
again for $r>2\,M$.
We also note that asymptotic flatness at $r\to\infty$ requires $K\le-1$, with
$K=-1$ yielding the pure Schwarzschild metric.
\par
Next, we note that the effective density is given by
\begin{equation}
k^2\,\tilde{\rho}
=
\frac{\alpha}{K\,r^2}
\left(\frac{\ell_p}{r}\right)^{1+1/K}
\left(1-\frac{2\,M}{r}\right)^{-1-1/K}
\ ,
\label{efecdenP}
\end{equation}
and diverges at $r=2\,M$ unless $-1<K<0$.
Of course, the effective pressure $\tilde{p}_{r}=K\,\tilde{\rho}$ also diverges at $r=2\,M$ unless $-1<K<0$.
The effective tangential pressure is given by
\begin{eqnarray}
k^2\,\tilde{p}_{t}
=
-\alpha\,\theta_2^{\ 2}
&=&
-\frac{\alpha\,(K+1)}{2\,K\,r^2}\left(1-\frac{M}{r}\right)
\left(\frac{\ell_p}{r}\right)^{1+1/K}
\nonumber\\
&&\left(1-\frac{2\,M}{r}\right)^{-2-1/K}
\ ,
\label{efecptanCP}
\end{eqnarray}
which also diverges at at $r=2\,M$ unless $-1/2<K<0$.
To summarise, the surface $r=\rH$ is a real singularity unless $-1/2<K<0$.
However, this is not compatible with the asymptotically flat conditions, which requires $K\le -1$.
We therefore conclude that the Killing horizon at $r=\rH=2\,M$ has become a real singularity,
which is not hidden inside a larger horizon.
\subsection{Linear equation of state}
Now let us consider a generic equation of state in the form
\begin{equation}
\label{generic}
\theta_0^{\,0}
=
a\,\theta_1^{\,1}+b\,\theta_2^{\,2}
\ ,
\end{equation}
with $a$ and $b$ constants.
The conformal case of Section~\ref{cs} is represented by the set $a=-1$ and $b=-2$, whereas the
polytropic $\Gamma=1$ (barotropic) case of Section~\ref{pc} is given by $a=-1/K$ and $b=0$.
Eqs.~(\ref{ec1d})-(\ref{ec3d}) then yield the differential equation for the MGD function
\begin{eqnarray}
\label{giso}
&&f^{*'}\left[\frac{1}{r}-\frac{b}{4}\left(\xi'+\frac{2}{r}\right)\right]
\nonumber\\
&&+ f^{*}\left[\frac{1}{r^2}-a\left(\frac{1}{r^2}+\frac{\xi'}{r}\right)
-\frac{b}{4}\left(2\,\xi''+\xi'^2+2\frac{\xi'}{r}\right)\right]
=
0
\ ,
\nonumber\\
\end{eqnarray}
whose general solution for $r>\rH=2\,M$ is given by
\begin{equation}
f^*(r)
=
\left(1-\frac{2\,M}{r}\right)
\left(\frac{\ell}{r-B\,M}\right)^{A}
\ ,
\label{giso2}
\end{equation}
where $\ell$ is a positive constant with dimensions of a length, and
\begin{eqnarray}
\label{AB}
A
&\!\!=\!\!&
\frac{2\,(a-1)}{b-2}>0
\\
B
&\!\!=\!\!&
\frac{b-4}{b-2}
\ ,
\end{eqnarray}
with $b\neq 2$ and the condition $A>0$ required by asymptotic flatness.
Therefore the solution reads
\begin{equation}
\label{Gsol}
e^{-\lambda}
=
\left(1-\frac{2\,M}{r}\right)
\left[1+\alpha
\left(\frac{\ell}{r-B\,M}\right)^{A}\right]
\ ,
\end{equation}
which again shows the horizon at $\rH=2\,M$, beside a possible divergence at $r=r_{\rm c}$
and a second zero at $r=r_0$, like in the previous cases.
\par
The physical content of the system is again clarified by the explicit computation of the effective
density 
\begin{equation}
\tilde{\rho}
=
\alpha\,\theta_0^{\ 0}
=
-\frac{\alpha}{k^2\,r^2}
\left(\frac{\ell}{r-B\,M}\right)^A
\left[1-A\left(\frac{r-2\,M}{r-B\,M}\right)\right]\ ,
\label{Gefecden}
\end{equation}
the effective radial pressure
\begin{equation}
\tilde{p}_{r}
=
-\alpha\,\theta_1^{\ 1}
=
\frac{\alpha}{k^2\,r^2}
\left(\frac{\ell}{r-B\,M}\right)^A
\ ,
\label{Gefecprera}
\end{equation}
and the effective tangential pressure
\begin{equation}
\tilde{p}_{t}
=
-\alpha\,\theta_2^{\ 2}
=
-\frac{\alpha\,A}{2\,k^2\,r^2\,\ell}
\left(\frac{\ell}{r-B\,M}\right)^{A+1}
\left(r-M\right)
\ .
\label{Gefecptan}
\end{equation}
Again, we see that the effective density and effective pressures diverge at 
\begin{equation}
\label{diver}
r_{\rm c}
=
B\,M
\ ,
\end{equation}
which represents a true singularity at $0<r_{\rm c}<\rH$ for $0<B<2$, that is
for
\begin{equation}
b<0
\qquad
{\rm or}
\quad
b>4
\ .
\end{equation}
For $B>2$ (that is, $0<b<2$), this singularity occurs outside the Killing horizon,
$r_{\rm c}>\rH$, and this case cannot be considered any further.
Secondly, the effective density and effective pressures satisfy
\begin{equation}
\tilde{p}_t
=
-\frac{1}{2}\left(\frac{r-M}{r-2\,M}\right)
\left(\tilde{\rho}+\tilde{p}_r\right)
\ ,
\end{equation}
showing that $\tilde{p}_t<0$ when both $\tilde{\rho}$ and $\tilde{p}_r$ are positive.
We thus conclude that at least one of the thermodynamic variables will always be negative
as long as the equation of state is linear.
Moreover, the effective radial and tangential pressure are related by
\begin{equation}
\label{rel}
\tilde{p}_t
=
-\frac{A}{2}\left(\frac{r-M}{r-B\,M}\right)
\tilde{p}_r
\ .
\end{equation}
Since $A>0$, we conclude that the two pressures always have opposite signs and
one of them will be negative.
On the other hand, the effective density and effective radial pressure are related by
\begin{equation}
\label{rel2}
\tilde{\rho}
=
\left[A\left(\frac{r-2\,M}{r-B\,M}\right)-1\right]
\tilde{p}_r
\ ,
\end{equation}
hence 
\begin{equation}
\label{rel3}
\tilde{\rho}\sim
\left\{
\begin{array}{ll}
-\tilde{p}_r
&
{\rm for}\
r\sim\,2\,M
\\
\left(A-1\right)\,\tilde{p}_r
&
{\rm for}\ 
r\gg 2\,M
\ .
\end{array}
\right.
\end{equation}
Since $A>0$, the asymptotic behaviour in Eq.~\eqref{rel3} demands $0<A\leq 1$ in order to ensure
that the density does not change its sign in the region $2\,M<r<\infty$
[the pressure~\eqref{Gefecprera} always has the same sign inside this region].
We conclude that the dominant energy condition $\tilde{\rho}\geq\mid\tilde{p}_r\mid$ cannot be satisfied
with a linear equation of state of the form displayed in Eq.~\eqref{generic}.
Nonetheless, the effective density is positive everywhere if $\alpha<0$~\footnote{Notice that for the particular
case $B=2$, namely the barotropic fluid, the density does not change its sign.
This remains an interesting exterior solution for a self-gravitating system of radius $R>\rH$.}.
\par
For $2<b<4$ one has $r_{\rm c}<0$ and there is no extra singularity beside the usual Schwarzschild one at $r=0$.
In this case, we must demand that no second zero $r_0>0$ of $e^{-\lambda}$ exists, otherwise the space-time
signature would become unacceptable inside a portion of $r>0$.
This condition is immediately satisfied if $\alpha>0$, for any $A>0$, that is for $a>1$.
We next notice that there is a second zero of $e^{-\lambda}$ at
\begin{equation}
\label{Gsz}
r_0
=
B\,M+{\ell}\,(-\alpha)^{1/A}
>
r_{\rm c}
\ ,
\end{equation}
when $\alpha<0$.
To have a proper black hole solution, this second zero $r_0$ must lie inside the relevant singularity.
If $2<b<4$, the relevant singularity occurs at $r=0$ an we must have 
\begin{equation}
\label{proper}
r_0
\le
0
\ ,
\end{equation}
that is, if $\ell$ and $|\alpha|$ are small enough to satisfy
\begin{equation}
{\ell}\,(-\alpha)^{1/A}
\le
-B\,M
\ .
\label{clinL}
\end{equation}
Otherwise, if $b<0$ or $b>4$, the relevant singularity occurs at $0<r_{\rm c}<\rH$, but 
$r_0>r_{\rm c}$ makes this case unsuitable.
\par
The final conclusion is thus that the linear equation of state~\eqref{generic} always produces black holes
(with a Schwarzschild singularity at $r=0$) if $2<b<4$ and $a>1$, provided $\alpha>0$ or
$\alpha<0$ and Eq.~\eqref{clinL} holds.
\subsection{A particular solution with no extra singularity}
\label{fs}
\begin{figure}[t]
\center
\includegraphics[scale=0.28]{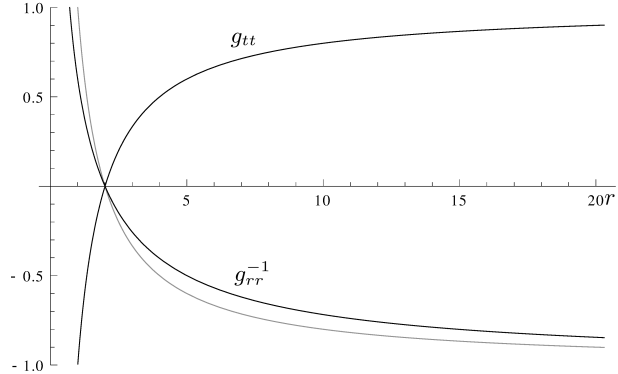}
\\
\centering\caption{Case $b = 3$.
Metric components for $\alpha=-0.7$ and $a=1.4$ (black lines) compared to the
Schwarzschild component $g^{-1}_{rr}$  (gray line).
The mass $M=1$.}  
\label{free}      
\end{figure}
The reader can see that Eq.~\eqref{generic} leads to a system very rich in possibilities,
whose generic solutions~\footnote{The case $b=2$ for the equation of state~\eqref{generic},
which is excluded in the solution~\eqref{Gsol}, yields a solution without any additional
singularity beside $r=0$, but with a switch in the sign of the density for $r>>M$.}
are given in Eqs.~\eqref{Gsol}-\eqref{Gefecptan}, and whose general analysis is detailed
throughout Eqs.~\eqref{diver}-\eqref{clinL}.
The main feature of these solutions is that they do not satisfy the dominant energy condition.
In this respect, let us recall that the energy conditions are a set of constraints which are usually
imposed on the energy-momentum tensor in order to avoid exotic matter sources,
hence they can be viewed as sensible guides to avoid unphysical situations.
However, it is well-known that these energy conditions might fail for particular classical systems
which are still reasonable~\cite{visser}.
In our case we are dealing with a gravitational source $\theta_{\mu\nu}$ whose main
characteristic is that it only interacts gravitationally with the matter that, by itself, would
source the (hairless) black hole solution~\eqref{Schw0}.
Hence, one should not exclude {\em a priori\/} that such matter is a kind of exotic source
(as indeed the conjectured dark matter is expected to be).
\par
Of all the possible solutions, we shall here analyse the particular case $b=3$ (with $a>1$
for asymptotic flatness) as an example of space-time which does not contain any extra
singularity beside the usual Schwarzschild one at $r=0$.
The radial metric component is obtained from Eq.~\eqref{Gsol} and reads
\begin{equation}
\label{Solb3}
e^{-\lambda}
=
\left(1-\frac{2\,M}{r}\right)
\left[1+\frac{\alpha}{(r+M)^{2\,(a-1)}}\right]
\ ,
\end{equation}
which makes it immediately clear that there is no second divergence.
In fact, the effective density is given by~\footnote{For simplicity we have redefined
$\alpha\,\ell^A\to\alpha$ in the right-hand side of Eqs.~\eqref{Solb3}-\eqref{b3efecptan}}
\begin{equation}
\tilde{\rho}
=
\alpha\,\theta_0^{\ 0}
=
\frac{\alpha}{k^2\,r^2}\left[\frac{2\,a\,(r-2\,M)-3\,(r-M)}{(r+M)^{2\,a-1}}\right]
\ ,
\label{b3efecden}
\end{equation}
the effective radial pressure by
\begin{equation}
\tilde{p}_{r}
=
-\alpha\,\theta_1^{\ 1}
=
\frac{\alpha}{k^2\,r^2\,(r+M)^{2\,(a-1)}}
\ ,
\label{b3efecprera}
\end{equation}
and the effective tangential pressure by
\begin{equation}
\tilde{p}_{t}
=
-\alpha\,\theta_2^{\ 2}
=
-\frac{\alpha\,(a-1)\,(r-M)}{k^2\,r^2\,(r+M)^{2\,(a-1)}}
\ .
\label{b3efecptan}
\end{equation}
The reader can easily check that the deformed Schwarzschild metric~\eqref{Schw} with
$g^{rr}=e^{-\lambda}$ in Eq.~\eqref{Solb3}, along with the source terms in
Eqs.~\eqref{b3efecden}-\eqref{b3efecptan},
solve the complete Einstein equations~\eqref{ec1}-\eqref{ec3} with $\rho=p=0$. 
\par
\begin{figure}[t] 
\center
\includegraphics[scale=0.29]{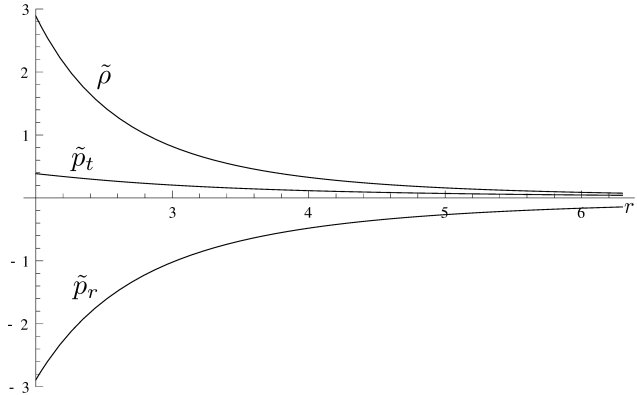}
\\
\centering\caption{Case $b = 3$. 
Effective source terms $\{\tilde{\rho},\,\tilde{p}_r,\,\tilde{p}_t\}\times 10^3$
for $\alpha=-0.7$ and $a=1.4$.
The horizon $\rH=2\,M$ and the mass $M=1$.}  
\label{free2}      
\end{figure}

Combining the expressions~\eqref{b3efecden} and~\eqref{b3efecprera}, we get 
\begin{equation}
\label{useful}
\tilde{\rho}=\left[\frac{2\,a\,(r-2\,M)-3\,(r-M)}{(r+M)}\right]\tilde{p}_r\ ,
\end{equation} 
from which we obtain the asymptotic behaviours
\begin{equation}
\label{near-far}
\tilde{\rho}\sim
\left\{
\begin{array}{ll}
-\tilde{p}_r
&
{\rm for}\
r\sim\,2\,M
\\
\\
(2\,a-3)\,\tilde{p}_r
&
{\rm for}\ 
r\gg 2\,M
\ .
\end{array}
\right.
\end{equation}
Therefore, when $\alpha<0$, the pressure $\tilde{p}_r<0$ and the effective energy will always
be positive for $r>2\,M$ whenever $a\leq\,3/2$.
Figs.~\ref{free} and~\ref{free2} show the corresponding metric elements and density and pressures
in~\eqref{b3efecden}-\eqref{b3efecptan} respectively for $\alpha=-0.7$ and $a=1.4$.
%
%
%
%
%
%
%
%
%
\section{Conclusions}
\label{s6}
\setcounter{equation}{0}
By making use of the MGD-decoupling approach, we have presented in detail how the Schwarzschild
black hole is modified when the vacuum is filled by a generic spherically symmetric gravitational fluid,
described by a ``tensor-vacuum'' $\theta_{\mu\nu}$, which does not exchange energy-momentum
with the central source.
For this purpose, we have separated the Einstein field equations into i) the Einstein equations for the spherically
symmetric vacuum $\rho=p=0$ and ii) the ``quasi-Einstein" system in Eqs.~(\ref{ec1d})-(\ref{ec3d})
for the spherically symmetric ``tensor-vacuum'' $\theta_{\mu\nu}$.
Following the MGD procedure, the superposition of the Schwarzschild solution found in i) plus the
solution for the ``quasi-Einstein" system in ii), has led to the solution for the complete system
``Schwarzschild + tensor-vacuum.''
\par
The quasi-Einstein system~(\ref{ec1d})-(\ref{ec3d}) was solved by providing some physically motivated
equations of state for the source $\theta_{\mu\nu}$.
In this respect, four different scenarios were considered, namely, i) the isotropic $\theta_1^{\,1}=\theta_2^{\,2}$;
ii) the conformal $\theta_\mu^{\,\,\mu}=0$; iii) the polytropic $\alpha\,\theta_1^{\,1}=K\,(\alpha\,\theta_0^{\,0})^\Gamma$
and iv) the generic linear equation of state in~\eqref{generic}.
In the isotropic case, we only found a metric which is not asymptotically flat for $r\to\infty$,
which means that the tensor-vacuum for a black hole cannot be isotropic as long as its interaction
with regular matter is purely gravitational.
On the other hand, the conformal case leads to the hairy black hole solution in Eq.~\eqref{confsol},
whose primary hairs is represented by the length $\ell$, which is constrained by the regularity
condition~\eqref{cscondition}.
Among all polytropic equations of state, we have only considered the barotropic $\Gamma=1$, which
represents a tensor-vacuum made of an isothermal self-gravitating sphere of gas.
This leads to the exterior solution in Eq.~\eqref{poly} endowed with the parameters $\{M,\alpha,\ell_{p},K\}$.
Since the Killing horizon $r=\rH=2\,M$ becomes a real singularity, this solution may represent the exterior
of a self-gravitating system of mass $M$ and radius $R>\rH$ but not a black hole solution.
\par
Finally, we have analysed the generic linear equation of state in Eq~\eqref{generic}, which includes
both the conformal and barotropic fluids as particular cases.
This leads to the solution in Eq.~\eqref{Gsol}, showing that even a simple linear equation of state may
yield hairy black hole solutions with a rich geometry described by the parameters $\{M,\alpha,\ell,a,b\}$,
where $\{\alpha,\ell,a,b\}$ represents a potential set of charges generating primary hairs.
In this context, a particular black hole solution with primary hairs $\{\alpha,a\}$ was found in Eq.~\eqref{Solb3},
whose main characteristic is the absence of other singularities in the region $0<r<\infty$.
\par
All the black holes solutions mentioned above have the horizon at $r_{\rm H}=2\,M$ and primary hairs
represented by a number of free parameters.
However, these parameters can be restricted by demanding i) the correct asymptotic behaviour and
ii) regularity conditions for black hole solutions free of pathologies.
In this respect, there are always a potential singularity $r_{\rm c}$ and a possible second horizon $r_{\rm h}$
in our solutions.
In order to have a proper black hole, it is necessary that $r_{\rm c}\leq\rH$ to avoid a naked singularity,
and $r_{\rm h}=\rH$ to have a metric with a proper signature.
We emphasize that $r_{\rm h}>\rH$ yields both $g_{tt}$ and $g_{rr}$ positive inside the region $\rH<r<r_{\rm h}$.
All these conditions yields restrictions on potential primary hairs.
For instance, the linear equation of state~\eqref{generic} always produces black holes if $2<b<4$ and $a>1$,
provided $\alpha>0$ or $\alpha<0$ and Eq.~\eqref{clinL} holds.
\par
We have shown that different characteristics of the gravitational source lead to different hairy black hole solutions.
Therefore, the compatibility between some of these solutions and the observations could determine the main features
of the tensor-vacuum, and eventually the fundamental field(s) that constitute it.
Finally, we would like to emphasize that the non-existence of an isotropic tensor-vacuum that does not
exchange energy-momentum with regular matter favours scenarios with Klein-Gordon type fields $\phi$,
which naturally induce anisotropy in the Einstein field equations.
These scalar fields are found in a large number of alternative theories to general relativity.
\section{Acknowledgements}
\par
J.O.~and S.Z.~have been supported by the Albert Einstein Centre for Gravitation and Astrophysics financed
by the Czech Science Agency Grant No.14-37086G.
R.C.~is partially supported by the INFN grant FLAG and his work has been carried out in the framework
of GNFM and INdAM and the COST action {\em Cantata\/}.
R.dR.~is grateful to CNPq (Grant No. 303293/2015-2), and to FAPESP (Grant No. 2017/18897-8)
for partial financial support.
A.S.~is partially supported by Project Fondecyt 1161192, Chile.
%

%
%


\begin{thebibliography}{99}
\bibitem{com}  This represents the simplest and so far the only known way to decoupling both gravitational sources in~\eqref{coupling}. An extension of the MGD approach (which represents the foundation of the MGD-decoupling) where both metric components are deformed, was developed in Ref.~\cite{MGDextended1}, but it works only in the vacuum and fails for regions where matter is present, since the Bianchi identities are no longer satisfied.
\end{thebibliography}
\end{document}